\documentstyle[aps,twocolumn]{revtex}
\begin{document}
\title{
Folded modes in the infrared spectra \\
of the spin-Peierls phase of CuGeO$_3$
}
\author{M.N.Popova, A.B.Sushkov, and S.A.Golubchik }
\address{Institute of Spectroscopy of Russian Academy of Sciences,
142092 Troitsk, Moscow
reg., Russia}
\author{A.N.Vasil'ev and L.I.Leonyuk}
\address{Moscow State University, 119899 Moscow, Russia}
\date{\today}
\maketitle
\begin{abstract}
Polarized far-infrared transmittance spectra of CuGeO$_3$ single
crystals were measured at different temperatures (6\,K$< T <$ 300\,K).
Two spectral lines,
at 284.2\,cm$^{-1}$ in ${\bf E}||{\bf c}$ polarization and
at 311.7\,cm$^{-1}$ in ${\bf E}||{\bf b}$ polarization,
appear at the temperature of the spin-Peierls transition and grow in
intensity with decreasing temperature. We assign these spectral
features to the folded modes of the dimerized lattice. We discuss a
possible role of the spin-phonon interaction in the formation of the
311.7\,cm$^{-1}$ feature.
\end{abstract}

CuGeO$_3$ has attracted a considerable attention as the first
inorganic compound that undergoes the spin-Peierls transition \cite{c1}.
The spin-Peierls transition is known to occur in
quasi-one-dimensional s=1/2 Heisenberg antiferromagnets due to
the magnetoelastic coupling between magnetic chains and a
three-dimensional phonon field. As a result of this coupling,
magnetic atoms displace along the chain and form nonmagnetic
dimers yielding a singlet ground state and a triplet excited state at
an energy $\Delta$ (the spin-Peierls gap).

In the high-temperature phase of
CuGeO$_3$ (space group $Pbmm$, $z$=2, lattice parameters $a$=0.480\,nm,
$b$=0.847\,nm, and $c$=0.294\,nm \cite{B}),
S=1/2 Cu$^{2+}$ ions occupy $C_{2h}$ symmetry positions in the
centra of strongly deformed edge-sharing Cu(O2)$_4$(O1)$_2$ octahedra
that form chains along the $c$-axis of the crystal (see Fig.\,\ref{f1}).
The octahedra from the neighboring chains share common O1
apical oxygens. As the Cu--O1  distances (0.275\,nm) are
appreciably greater than the Cu--O2 distances (0.193\,nm),
well isolated ribbons of Cu(O2)$_4$ rectangles running along the
$c$-axis may be considered as main magnetic units.

Below $T_{\rm sp}\approx$14\,K the Cu$^{2+}$ chains distort into dimers
and the spin-Peierls gap $\Delta_1$=2\,meV=16\,cm$^{-1}$ opens at the
point ${\bf k}_{\rm AF}$=(0, 1, 1/2) of the Brillouin zone
(BZ)~\cite{c4,R-A}.
The narrowest gap is shifted from the ${\bf k}=0$ point because of
the interchain interaction.
A second gap $\Delta_2$=5.5\,meV=44.3\,cm$^{-1}$ has been found in the
electron spin resonance \cite{Brill} and far infrared (FIR)
absorption~\cite{L} experiments and has been assigned to the point
${\bf k}$=(0, 0, 0). In a recent theoretical paper \cite{U} it was,
however, suggested to place this gap at the point ${\bf k}$=(0, 1, 0).
X-ray, neutron and electron diffraction studies revealed superlattice
reflections that have been indexed with a commensurate
propagation vector ${\bf k_{\rm sp}}$=(1/2, 0, 1/2) corresponding 
to a crystal unit cell doubled along the $c$- and $a$-directions 
(see, e. g., \cite{B} and references therein).
This doubling of the unit cell leads to a folding of the Brillouin
zone and, as a result, to a transfer of phonons from the zone
boundary to its center.
The structure of the spin-Peierls phase belongs to the $Bbcm$ space
group \cite{B}.
The factor group analysis shows that additional eighteen Raman active
modes $4A_g+5B_{1g}+4B_{2g}+5B_{3g}$
and nine IR active modes
$2B_{1u}({\bf E}||{\bf c})+4B_{2u}({\bf E}||{\bf b})+3B_{3u}({\bf E}||{\bf a})$
have to appear in the spin-Peierls phase of CuGeO$_3$.
(Here, symmetries of the vibrational modes are given according to the
axes setting ${\bf x}||{\bf a}, {\bf y}||{\bf b}, {\bf z}||{\bf c}$).

The Raman peaks at 369 and 820\,cm$^{-1}$ have been attributed to the
$A_g$ folded modes \cite{c8,c10,M}.
The feature at 43.5\,cm$^{-1}$ observed in the
far-infrared absorbance difference spectra
$\alpha(T)-\alpha$(15.5\,K) below $T_{\rm sp}$ has been supposed to
arise from the folded mode
corresponding to a TA phonon at the $a^*$ zone boundary \cite{L}.
The authors of Ref.\,\cite{L} could not investigate how does the
43.5\,cm$^{-1}$ feature behave when approaching $T_{\rm sp}$, because
of its proximity to the relatively strong 44.3\,cm$^{-1}$ peak due to
the $\Delta_2$ spin gap.
The 311.7\,cm$^{-1}$ feature observed in both transmittance
$T_r$(20\,K)/$T_r$(6\,K) \cite{MW} and reflectance
$R$(20\,K)/$R$(5\,K)      \cite{c12} ratios for CuGeO$_3$ has been
ascribed to a softening of the $B_{2u}$ ($B_{3u}$ in $Pmma$ setting)
phonon mode near 300\,cm$^{-1}$ when passing through the 14\,K
spin-Peierls transition to the dimerized state \cite{c12}.
All the above mentioned IR measurements have been performed using
unpolarized light and relatively thick samples with saturated
absorption by phonons.

Recently, we have shown that lattice dimerization in the course of
the spin-Peierls transition in NaV$_2$O$_5$, in contrast to the case
of CuGeO$_3$, leads to the appearance of numerous folded modes in the
FIR transmittance spectra with typical integrated intensities
$\int d\omega\,\alpha\approx 100-400$\,cm$^{-2}$  \cite{nav1}.
In attempt to clarify the nature of such a puzzling difference
between these two recently discovered inorganic spin-Peierls
compounds, we have undertaken the present study.
This paper reports on a direct measurement of the polarized FIR
transmittance spectra of CuGeO$_3$ as a function of temperature.
In the spectral range studied (60--400\,cm$^{-1}$), we clearly
observe the appearance of two absorption lines (at 284.2 and
311.7\,cm$^{-1}$) at the temperature of the spin-Peierls transition.
We attribute these spectral features to the folded modes which occur
due to lattice dimerization in the spin-Peierls phase and speculate
that the spin-phonon interaction contributes to the 311.7\,cm$^{-1}$
absorption line.

Blue in color transparent single crystals of CuGeO$_3$ were
grown from the melt of high purity oxides by a method of
spontaneous crystallization under a slow cooling. They were
checked with X-ray diffraction and magnetization measurements
and exhibited the temperature of the spin-Peierls transition
$T_{\rm sp}=14.0\pm0.2$\,K.
Three samples cleaved perpendicular to the $a$-axis and having
dimensions approximately 3$\times$3\,mm along the $b$- and 
$c$-axes were used in our experiments.
Their thicknesses were 240, 20$\pm 1$ and 0.7$\pm 0.1$\,$\mu$m. 
The thinnest
sample was on a Scotch tape.  Samples were mounted in a special
insert that was put into a variable temperature ($T$=5--300\,K)
helium vapor optical cryostat. Polarized infrared transmission
measurements were performed in the frequency range from 60 to
400\,cm$^{-1}$ with the spectral resolution 0.2--1\,cm$^{-1}$, using
a BOMEM DA3.002 Fourier transform spectrometer. To avoid possible
errors connected with a thermal displacement of a sample relative to
the radiation beam within the spectrometer we measured reference
spectra at each temperature. Our insert has been specially
constructed for such a procedure. In our experimental geometry
(${\bf k}||{\bf a}$) we could anticipate to find
four new $B_{2u}$ modes in ${\bf E}||{\bf b}$ polarization and
two  new $B_{1u}$ modes in ${\bf E}||{\bf c}$ polarization below $T_{\rm sp}$.

Fig.\,\ref{f2}a shows the transmittance spectrum
in ${\bf E}||{\bf b}$
polarization of the 20\,$\mu$m thick sample at temperatures above
and below $T_{\rm sp}$. Two well known \cite{c13} $B_{2u}$ phonon
modes near 220 and 300\,cm$^{-1}$ are seen in the displayed spectral
range. From the intervals of zero transmittance we determine the
corresponding transverse and longitudinal frequencies. We find
$\omega_{\rm TO}$=210 and 284\,cm$^{-1}$, $\omega_{\rm LO}$=233 and
310 cm$^{-1}$ at room temperature, in good agreement with parameters
found from the best oscillator-fit of reflectivity spectra
\cite{c13}. The structure with regularly changing period in Fig.\,\ref{f2}a
comes from the interference in the platelet, taking into account the
changes of the refractive index in the vicinity of a phonon
frequency. We have confirmed such an interpretation by direct
calculations of transmittance spectra of a platelet with a given
thickness and the dielectric constant taken from the
Ref.\,\cite{c13}. The main change in the spectrum when passing
through the spin-Peierls transition temperature is a blue shift of
the high frequency edge of the phonon mode near 300\,cm$^{-1}$.

To investigate this phenomenon, we have measured 
the 0.7\,$\mu$m thick sample.
Two sharp peaks at the $\omega_{TO}$ frequencies are present in its
spectrum (see the spectrum $b$ in Fig.\,\ref{f2}). 
It is clearly seen that a
new spectral line at 311.7\,cm$^{-1}$ appears below $T_{\rm sp}$.
Fig.\,\ref{f3}a demonstrates how does this line grow in intensity
with decreasing temperature. We did not observe any softening of the
288\,cm$^{-1}$ phonon mode near $T_{\rm sp}$. This mode smoothly
shifts to higher frequencies when going from the room temperature
down to 6\,K.

In the ${\bf E}||{\bf c}$ spectrum of the uniform phase, only one
phonon mode is present (at 165\,cm$^{-1}$), and it is narrow
($\omega_{\rm LO}-\omega_{\rm TO}$=3\,cm$^{-1}$). So, we were able to
measure transmittance of thick samples. The spectra of the 240\,$\mu$m
thick sample revealed a new spectral line at 284.2\,cm$^{-1}$ that
appeared below $T_{\rm sp}$ (see Fig.\,\ref{f3}b).

Table\,1 lists the parameters of the two observed lines. For both of
them, the position and the full width at half height (FWHH)
do not depend on the temperature, within the experimental precision.
The normalized integrated intensity versus temperature
dependences are practically
identical for both lines (see the results for the 240\,$\mu$m and
20\,$\mu$m thick samples presented in Fig.\,\ref{f5}). The shift of the
transition temperature to $T_{\rm sp}$=15.2\,K for the thinnest sample
may be due to internal stress within this 0.7\,$\mu$m thick platelet
on a Scotch tape (it is known that $T_{\rm sp}$ becomes higher under
pressure \cite{c14}).  The same might be the reason for a broader
line in this sample as compared to the 20\,$\mu$m thick sample.
At $T>T_{\rm sp}$ both lines are absent, within the precision of our
measurements which may be estimated as 10\,\% for the
284.2\,cm$^{-1}$ line and 5\,\% for the 311.7\,cm$^{-1}$ one,
taking into account the signal-to-noise ratio in our spectra and the
measurements of the samples of different thickness.

Let us discuss the possible nature of the observed FIR
absorption lines. We shall consider purely magnetic, purely phonon
and mixed absorption which could occur in the spin-Peierls phase of
CuGeO$_3$. In this phase, well defined magnonlike triplet excitations
have been observed by  inelastic neutron scattering (INS) \cite{c4}.
Their frequencies show a noticeable dispersion
in the $b^{*}$ and $c^{*}$ directions of the BZ.
A singlet-triplet transition has been observed as a FIR absorption
peak at 44.3\,cm$^{-1}$ that splits in a magnetic field \cite{L}.
Absorption with creation of two magnons (${\bf k}$ and
$-{\bf k}$) would reflect a magnon density of states.
Then, one could expect spectral features at the doubled
frequencies corresponding to flat area in the dependence
$\omega({\bf k})$, namely, at
$2\Delta_1=32$\,cm$^{-1}$,
$2\Delta_2=90$\,cm$^{-1}$, $2\omega_m=255\pm 7$\,cm$^{-1}$ \cite{c4,R-A},
$\omega_m$ corresponding to the maximum of the dispersion curve.
A possible magnon-magnon interaction would somewhat lower these frequencies.
The frequencies of the discussed FIR spectral lines, 284.2 and
311.7\,cm$^{-1}$, lie higher than the mentioned frequencies
which rules out their interpretation as
features due to two-magnon (2M) absorption. It should be mentioned
that for CuGeO$_3$ the most important term in the 2M
absorption Hamiltonian vanishes by symmetry considerations \cite{Moriya},
in contrast to the case of another
spin-Peierls compound, NaV$_2$O$_5$, where 2M absorption is
allowed. For NaV$_2$O$_5$, indeed, we succeeded to observe it \cite{nav1}.

Another absorption process that involves magnetic excitations is a
phonon-assisted two magnon (P+2M) absorption which may be particularly
important when 2M absorption is forbidden (see, e. g., \cite{Moriya}).
Critical points in the densities of phonon and magnon states
contribute to P+2M absorption peaks. Because of that, magnetic
excitations with energies $\Delta_1$, $\Delta_2$ and $\omega_m$ have
to be taken into consideration. All of them broaden dramatically with
increasing temperature. Besides, the gap modes shift to zero when
approaching $T_{\rm sp}$. If the observed spectral features (at 284.2 and
311.7\,cm$^{-1}$) were due to P+2M absorption they would shift and/or
broaden with increasing temperature towards $T_{\rm sp}$. This is not the
case, within the precision of our experimental data.

The absence of any broadening or shift of the observed FIR absorption
lines at 284.2 and 311.7\,cm$^{-1}$ with approaching T$_{\rm sp}$
from below rules out a participation of magnetic excitations in their
formation.
To our mind, the FIR absorption lines at 284.2 and 311.7\,cm$^{-1}$
result from zone boundary phonons activated by the zone
folding due to the lattice dimerization in the spin-Peierls phase.
Unfortunately, the
frequencies of zone boundary phonons are known
(from INS) only up to about 180\,cm$^{-1}$ \cite{N-F}, so, no
comparison with the frequencies of the observed FIR modes can be made.
The intensity of folded modes should be
proportional to squared displacements of atoms in the course of the
spin-Peierls transition, $I(T)\propto\delta^2(T)$, just as in the case
of superlattice reflections in X-ray or inelastic neutron scattering.
Our experimental $I(T)$ dependences displayed in Fig.\,\ref{f5} are similar
to those reported from INS \cite{R-A}, which is in agreement with the
interpretation of the peaks at 284.2 and 311.7\,cm$^{-1}$ as folded modes.

The peak at 311.7\,cm$^{-1}$ is much stronger and broader than the
one at 284.2\,cm$^{-1}$ (see Table~1).
We suppose, the reason for that lies in a relatively strong spin-phonon
coupling for the former mode.
Recent neutron diffraction experiments have shown that the
deformation of the lattice above $T_{\rm sp}$ may be characterized by a
rotation of the Cu(O2)$_4$ ribbons around the $c$-axis and has the
$A_g$ symmetry \cite{B}.
It is related to the distortion below $T_{\rm sp}$, in the sense that the
rotation of the O2 edges around the $c$-axis is the common element.
In the high-temperature deformation the edges stacked along the
$c$-axis are displaced in the same way, whereas below $T_{\rm sp}$
neighboring edges are displaced in opposite directions resulting in
the twist distortion. Both deformations strongly influence the
magnetic interaction parameter $J$ \cite{B}.
Keeping that in mind, we propose the zone boundary counterpart of the
$A_g$ mode at 332\,cm$^{-1}$ \cite{c13} as a candidate for a phonon
responsible for the observed 311.7\,cm$^{-1}$ feature.
At the zone center, this mode tends to rotate Cu(O2)$_4$ ribbons around the
$c$-axis \cite{c13} while at the zone boundary it corresponds to the
twist rotations of the neighboring O2--O2 edges.
It seems evident from the results of the work \cite{B} that this mode
efficiently modulates the superexchange interaction and, thus,
is strongly coupled to spins.
In the spin-Peierls phase $A_g$ mode splits into
$A_g+B_{3g}+B_{1u}+B_{2u}$ where the latter is indeed IR active in
${\bf E}||{\bf b}$ polarization.

In summary, our measurements of polarized FIR transmittance spectra
of CuGeO$_3$ have unambiguously shown that the $B_{2u}$ mode near
300\,cm$^{-1}$ does not soften upon passing through the 14\,K
spin-Peierls transition, as has been stated in Ref.\,\cite{c12}, but
a new mode with the frequency 311.7\,cm$^{-1}$ appears.
Another mode, of $B_{1u}$ symmetry, appears at 284.2\,cm$^{-1}$.
Both modes are, most likely, transferred to the zone center from the
zone boundary, due to lattice dimerization.
We propose a possible mechanism of the spin-phonon coupling
contributing to the 311.7\,cm$^{-1}$ mode.

We are grateful to G.N.Zhizhin for a support. This work was
made possible in part by Grants N95-02-03796-a and
N96-02-19474-a from the Russian Fund for Fundamental Research.

\newpage
\begin{figure}
\caption{Crystal structure of CuGeO$_3$.
The axes are shown in the $Pbmm$ setting.}
\label{f1}
\end{figure}

\begin{figure}
\caption{Transmittance spectra of (a) 20\,$\mu$m and (b) 0.7\,$\mu$m
thick samples of CuGeO$_3$ for ${\bf E}||{\bf b}$ polarization;
solid lines --- $T=6\,{\rm K} < T_{\rm sp}$, dashed line ---
$T=30\,{\rm K} > T_{\rm sp}$.}
\label{f2}
\end{figure}

\begin{figure}
\caption{FIR absorption lines activated by the spin-Peierls
transition in CuGeO$_3$:
(a) ${\bf E}||{\bf b}$, 0.7\,$\mu$m thick sample;
(b) ${\bf E}||{\bf c}$, 240\,$\mu$m thick sample.}
\label{f3}
\end{figure}

\begin{figure}
\caption{Temperature dependences of the integrated intensity
$I(T)=\int d\omega\,[\alpha(T,\omega)-\alpha(20\,K,\omega)]$
for the lines at 284.2\,cm$^{-1}$ (circles --- 240\,$\mu$m thick sample)
and 311.7\,cm$^{-1}$ (squares --- 20\,$\mu$m thick sample,
triangles --- 0.7\,$\mu$m thick one).}
\label{f5}
\end{figure}

\begin{table}[h]
\caption{Parameters at $T=6$\,K of the FIR absorption lines activated by
the spin-Peierls transition in CuGeO$_3$}

\begin{center}
\begin{tabular}{ l l l }

Polarization  &         ${\bf E}||{\bf c}$  & ${\bf E}||{\bf b}^{\dag}$ \\[2mm]
Frequency $\omega$, cm$^{-1}$  & 284.24$\pm$0.05 & 311.70$\pm$0.05  \\[2mm]
FWHH $\Delta\omega$, cm$^{-1}$ & 0.55$\pm$0.05  & 4.5$\pm$0.5$^{\ddag}$ \\[2mm]
Peak absorption &                         ~               & ~ \\
coefficient $\alpha_{\rm m}$, cm$^{-1}$ & 9.3$\pm$0.3     & 500$\pm$100 \\[2mm]
Integrated intensity,           &         ~               & ~ \\
cm$^{-2}$                       &         6$\pm$0.5       & 2500$\pm$500 \\

\end{tabular}
\end{center}
$^{\dag}$data for the 0.7\,$\mu$m thick sample. \\
$^{\ddag}$3.5$\pm$0.5 for the 20\,$\mu$m thick sample.
\end{table}

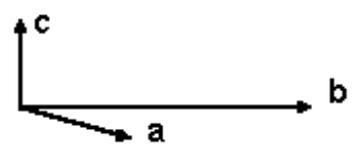
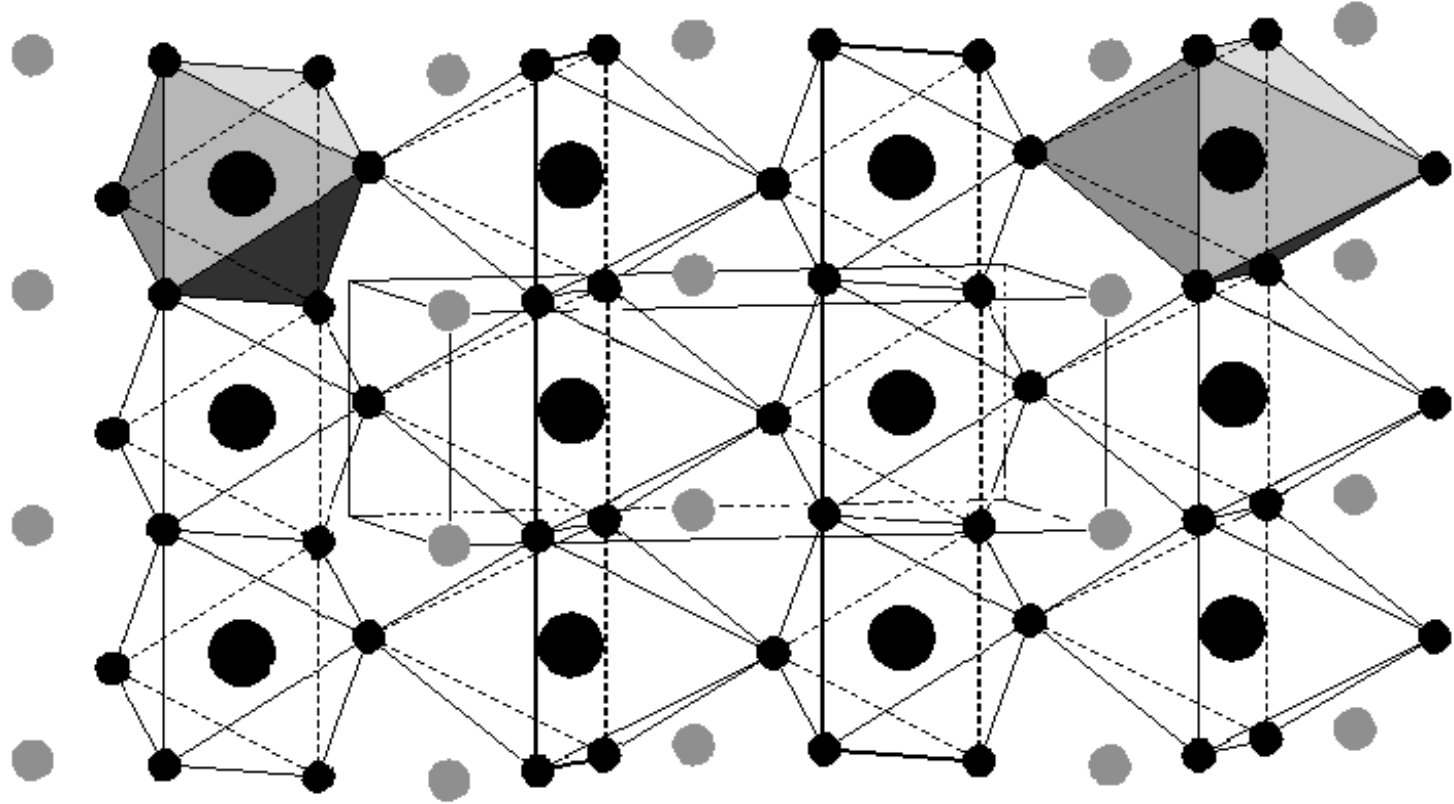
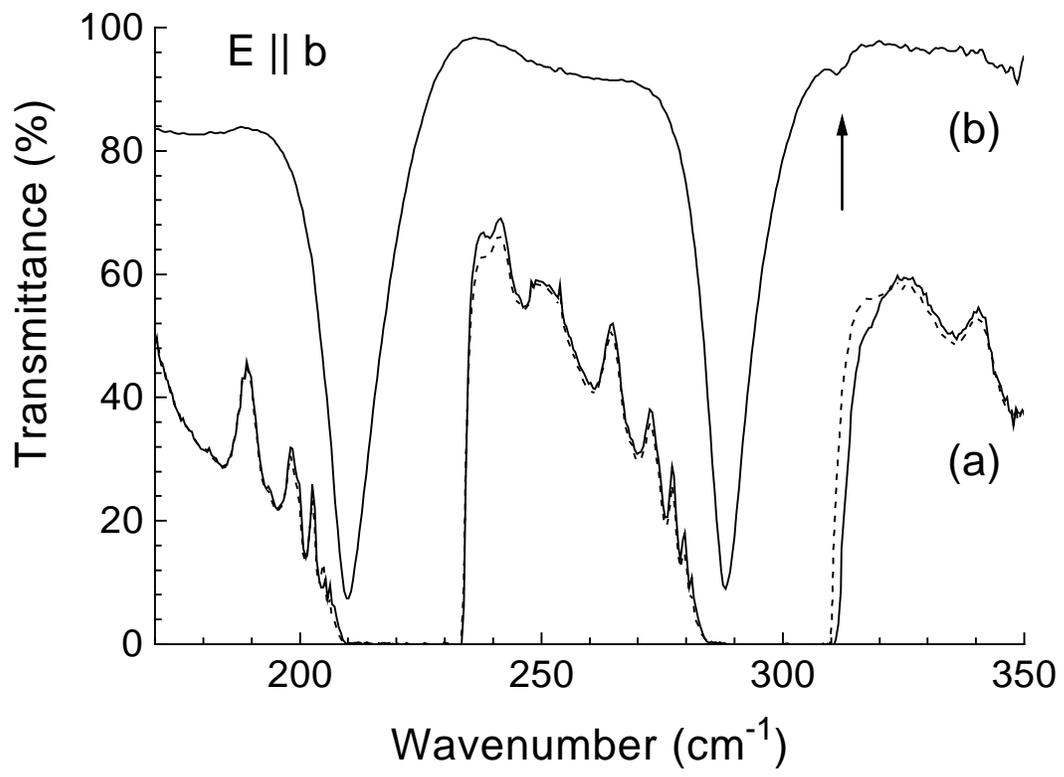
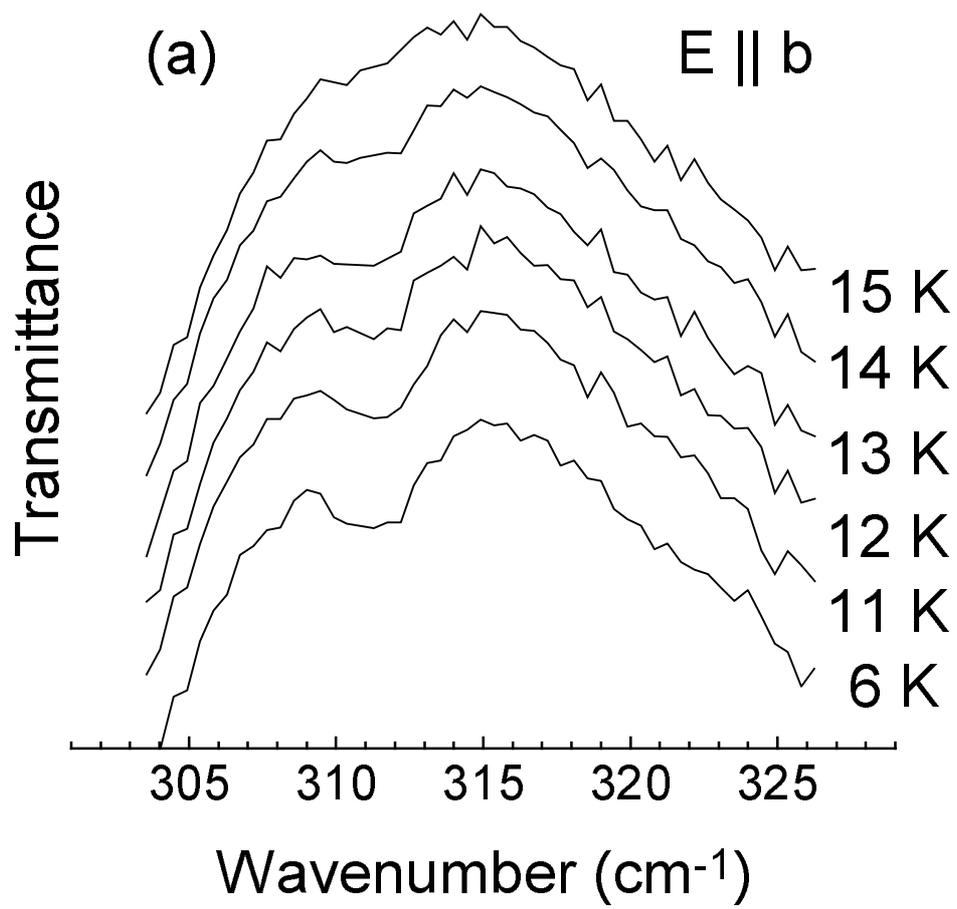
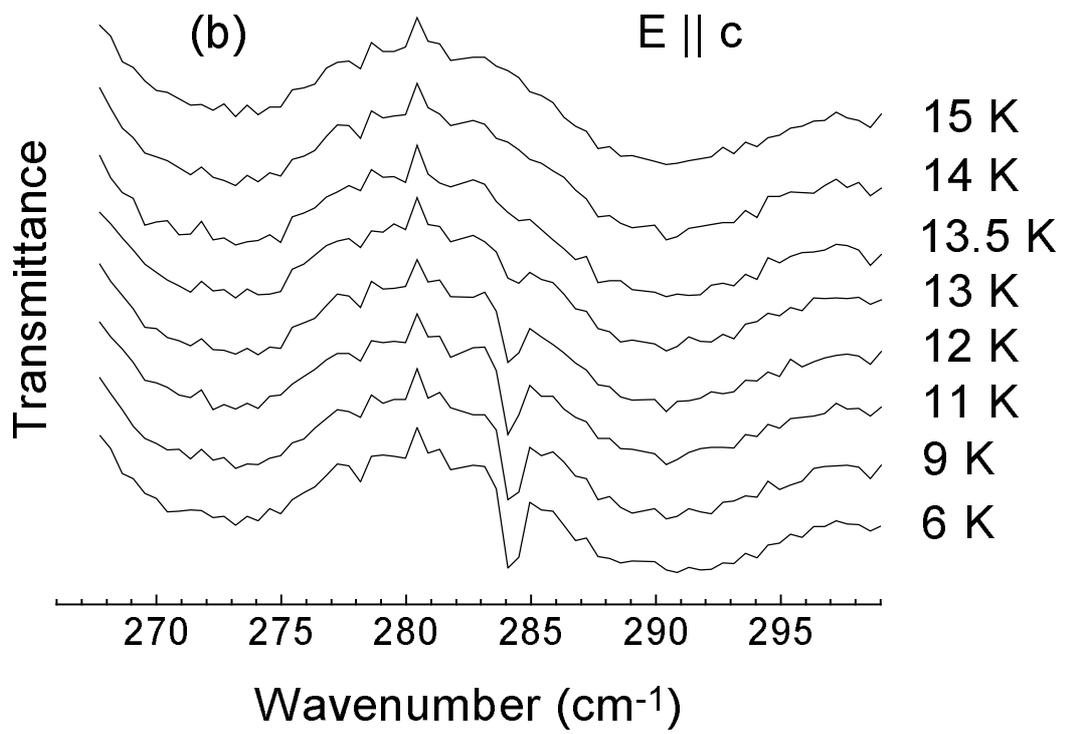
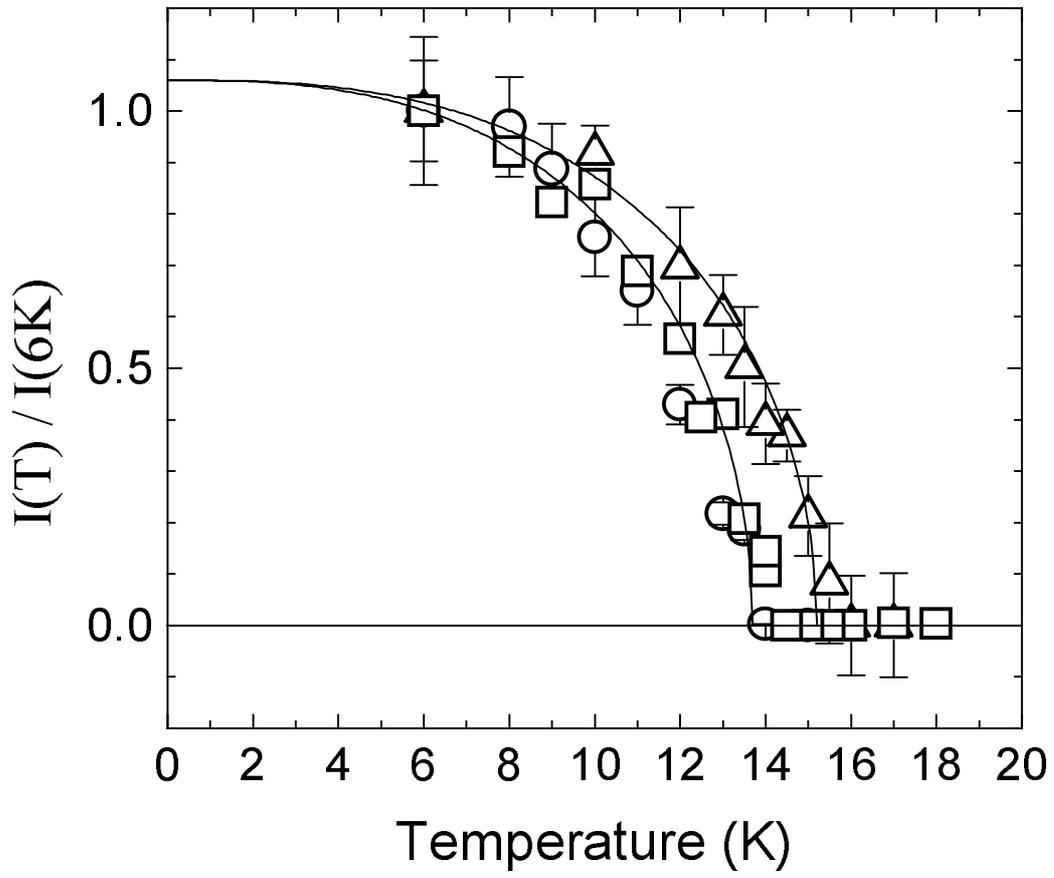
\end{document}